\documentclass[twocolumn]{revtex4}
\usepackage{amsmath, amsfonts, amssymb, graphicx, color}

\newcommand{\revision}{\color{black}}

\newcommand{\lastequal}{Corresponding authors. These authors contributed equally.}

\begin{document}

\newcommand{\deftitle}{{Learning the heterogeneous hypermutation landscape of immunoglobulins from high-throughput repertoire data}}

\title{\deftitle}

\author{Natanael Spisak}
\affiliation{Laboratoire de physique de l'\'Ecole normale sup\'erieure,
  CNRS, PSL University, Sorbonne Universit\'e, and Universit\'e de
  Paris, 75005 Paris, France}
\author{Aleksandra M. Walczak}
\thanks{\lastequal}
\affiliation{Laboratoire de physique de l'\'Ecole normale sup\'erieure,
  CNRS, PSL University, Sorbonne Universit\'e, and Universit\'e de
  Paris, 75005 Paris, France}
\author{Thierry Mora}
\thanks{\lastequal}
\affiliation{Laboratoire de physique de l'\'Ecole normale sup\'erieure,
  CNRS, PSL University, Sorbonne Universit\'e, and Universit\'e de
  Paris, 75005 Paris, France}

\begin{abstract}
Somatic hypermutations of immunoglobulin (Ig) genes occurring during affinity maturation drive B-cell receptors' ability to evolve strong binding to their antigenic targets. The landscape of these mutations is highly heterogeneous, with certain regions of the Ig gene being preferentially targeted. However, a rigorous quantification of this bias has been difficult because of phylogenetic correlations between sequences and the interference of selective forces. Here, we present an approach that corrects for these issues, and use it to learn a model of hypermutation preferences from a recently published large IgH repertoire dataset. The obtained model predicts mutation profiles accurately and in a reproducible way, including in the previously uncharacterized Complementarity Determining Region 3, revealing that both the sequence context of the mutation and its absolute position along the gene are important. In addition, we show that hypermutations occurring concomittantly along B-cell lineages tend to co-localize, suggesting a possible mechanism for accelerating affinity maturation.

\end{abstract}

\maketitle

\section{Introduction}

B cells are a crucial player in the adaptive immune system. Swift eradication of pathogens is enabled by the production of immunoglobulins (Ig) that bind tightly to antigens, helping in their detection, neutralization, and removal. Achieving high accuracy and breadth relies on the extraordinary diversity of the B cells repertoire. 
The process of V(D)J recombination results in a highly diverse population of naive cells \cite{Hozumi1976,Boyd2009a,Glanville2009,Harlan12,Elhanati2015,DeWitt2016,Briney2019}. In addition, B cells undergo affinity maturation, a Darwinian process \cite{Cobey2015} in which mutations are introduced to the immunoglobulin-coding gene and highest affinity mutants are selected \cite{Mesin2016}.
This process is driven by a very high rate of somatic hypermutations (SHM), $\sim10^{-3}$ per basepair per cell division \cite{Kleinstein2003}, targeting the Ig genes. Some receptor genes can ultimately accumulate up to 30\% amino acid substitutions, considerably altering the initial genotype. The broad diversity created by SHM ultimately ensures the emergence and selection of strong antigen binders. Understanding SHM and their statistics is key to designing better vaccination strategies \cite{Bonsignori2016,Schramm2018}.

Like the VDJ recombination process, SHM are characterized by heterogeneous preferences. Mutational pathways affect the Ig genes unevenly, with `cold' and `hot' spots along the receptor gene, even before somatic selection introduces further biases \cite{Schramm2018}. SHM is initiated by Activation-Induced cytidine Deaminase (AID) through the deamination of deoxycytidines triggering an array of error-prone repair pathways \cite{Feng2020}. AID and repair enzymes preferentially target certain regions of the gene. However, a quantitative picture of how these processes and their context dependencies result in the observed heterogeneous mutational landscape is lacking.
High-throughput repertoire sequencing of the Ig gene \cite{Weinstein2009,Boyd2009a,Glanville2009,Reddy2010} has facilitated the development of effective models from a detailed analysis of mutational profiles of Ig sequences before \cite{Yaari13,Elhanati2015,Cui2016} or after selection \cite{McCoy2015,Sheng2017,Hoehn2017,Dhar2018,Marcou2018}. However, the spatial organization of mutations, their context preferences, and their interplay with selection during affinity maturation are still poorly understood, in part due to a number of confounding factors.

A fundamental issue is the bias of selection, which favors beneficial mutations over deleterious ones in the observed repertoire. This bias can be partially circumvented by analyzing synonymous substitutions \cite{Yaari13}, with the limitation that extrapolation is required to generalize to non-synonymous ones.
Another way around selection is to study passenger nonproductive sequences, which are unsuccessful products of VDJ recombination and thus unaffected by selection \cite{Elhanati2015,Marcou2018,Cui2016}. These sequences make up a minority of DNA sequences, and are rarely found in mRNA sequences because of allelic exclusion, which limits their use to very large datasets. 

Another confounding factor arises from phylogenetic biases due to the complex multi-lineage structure of the repertoire. While methods have been developed to infer substitution rates from lineages in a lineage-specific \cite{Dhar2018} or repertoire-wide way \cite{Hoehn2019}, they do not aim to correct for selection and do not address the question of hypermutation targeting.

Here we propose a new framework for quantifying and predicting immunoglobulin mutability. The model is trained on the reconstructed phylogenies of nonproductive lineages from very large published B cell repertoires totalling around half a million nonproductive sequences \cite{Briney2019}, allowing us to overcome previous limitations of dataset sizes.
The approach accounts for both phylogenetic and selection biases, and allows us to study in detail the spatial and context preferences of hypermutation targeting, and to reveal the co-localization of contemporary mutations.

\section{Results}

\subsection*{Repertoire-wide framework to model intrinsic mutabilities from out-of-frame lineages}
Out-of-frame Ig sequences are byproducts of the VDJ recombination process that are made non functional by a frameshift in the CDR3 region. Since each cell has two copies of the Ig genes, out-of-frame rearrangements may survive in the cell if recombination on the second chromosome is successful. The mechanism of allelic exclusion ensures that only the functional variant is expressed. Yet, out-of-frame IgH sequences comprise $\sim 2\%$ of rearrangements in Ig mRNA sequencing experiments, and $\sim 9\%$ in genomic DNA \cite{DeWitt2016}. When a B-cell clone harboring both an out-of-frame and a functional rearrangement undergoes affinity maturation, the out-of-frame sequence acts as a passenger and mutates alongside the functional sequence, with the selection pressure acting only on the latter. While the two sequences share the same phylogeny, mutations found in out-of-frame lineages are not expected to be subject to selection.

\begin{figure*}
\begin{center}
\includegraphics[width=\textwidth]{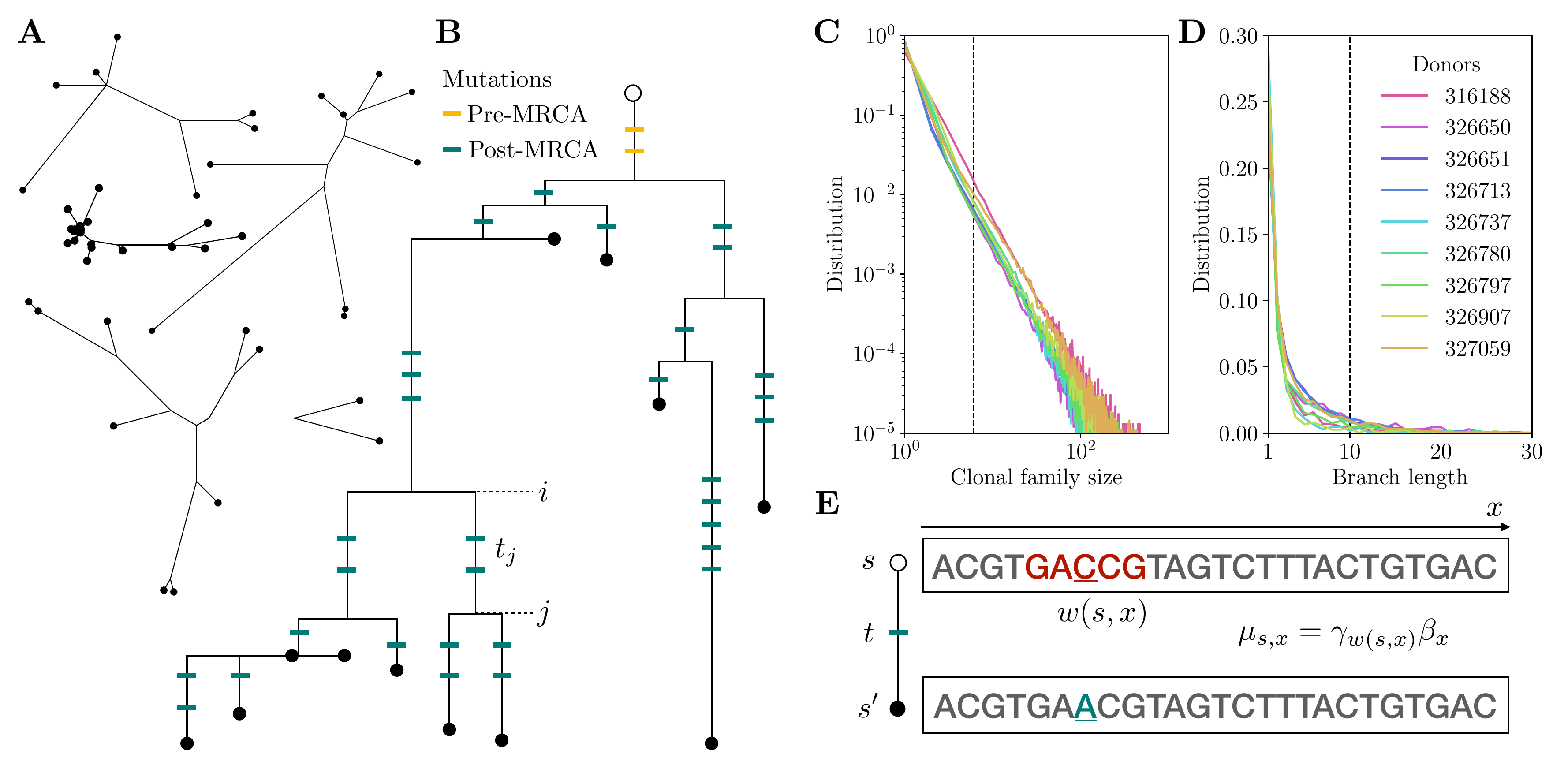}
\caption{Repertoire-wide framework to model somatic hypermutations (SHM) in out-of-frame lineages. A. Examples of out-of-frame clonal families. B. An example of a phylogeny with marked pre- and post-Most Recent Common Ancestor (MRCA) mutations. Only post-MRCA mutations are used for learning the somatic hypermutation model. C. Clonal family size distribution. Phylogenies were inferred for families with more than 5 unique sequences. D. Branch length distribution. Mutations encoded in branches shorter than 10 mutations were used for model inference. E. Context and position dependence of the mutation rate across a sequence. In this example the underlined C mutates to A, in the GACCG context denoted as $w(s,x)$. The mutation rate $\mu_{s,x}$ depends both on the sequence $s$ and the position $x$, through the context dependent rate $\gamma_{w(s,x)}$ and the position dependent rate $\beta_x$.}
\label{fig:cartoon}  
\end{center}
\end{figure*}

To model the process of SHM, we reconstruct the evolutionary history giving rise to the observed mutation patterns in nonproductive rearrangements. We analysed data consisting of the IgG repertoires of 9 individuals from Ref.~\cite{Briney2019}, obtained by the targeted mRNA sequencing of the Ig heavy (IgH) chain locus.
We pre-processed and aligned raw IgH sequences to keep only out-of-frame sequences. We then grouped sequences into clonal families that originate from the same ancestor using single linkage clustering (Fig.~\ref{fig:cartoon}A). The size of clonal families typically follows a power-law distribution (Fig.~\ref{fig:cartoon}C).
As a result, many lineages are represented by one or a few sequences.
We focused on sufficiently large lineages (comprised of at least 6 distinct sequences) and reconstructed their lineage structure, using maximum likelihood \cite{Felsenstein1981,Stamatakis2014} to infer the topology of the underlying tree, and marginal reconstruction for the identity of ancestral states. This provides us with a list $>200,000$ mutation events occurring between the most recent common ancestor of lineages and their leaves.

Using lineage information is essential for multiple reasons. First, it allows for a better estimate of the sequence context in which a mutation appears. In this paper we define the context as the 5-mer sequence  comprising the mutated basepair flanked by 2 basepairs on each side.
In the absence of lineage information, the best guess for the 5-mer context would be given by the genomic sequence of the V, D, or J segment where the mutation arose. But that context may itself be affected by other prior mutations. The tree structure allows us to identify the order of mutations and reconstruct the probable 5-mer context in which each mutation occurred. Second, for the same reason, the tree structure can help identify mutations in the hyper-variable CDR3 region, including in the junctions made of nontemplated insertions. This makes it possible to estimate the hypermutation rate in these regions. Together, these improvements mean that mutations can be identified within a broader range of 5-mer contexts, and their corresponding mutabilities better estimated. Third, lineage structure helps reduce contamination from sequences that have been under some selection. In some rare events, during affinity maturation a somatic insertion or deletion may be introduced in the CDR3 of a productive sequence, which would lead us to classify it as out-of-frame, even though it has been subject to selection prior to the frame-shift event. Focusing on mutations happening downstream of the most recent common ancestor, which is already out of frame, help us discards those contaminating events.

Given a model $P(s \rightarrow s' | t, \theta)$ of sequence evolution from $s$ to $s'$, where $t$ is fraction of mutated positions between $s$ and $s'$, (called branch length, equal to the number of mutations divided by alignment length), and $\theta$ denotes model parameters, we can write the joint likelihood of mutation events in each lineage as
\begin{equation}
\label{likelihood}
    P(S|T,\theta) = \prod_{(i,j)\in T} P(s_i \rightarrow s_j | t_j, \theta),
\end{equation}
where $S$ is the set of sequences (observed and reconstructed) at each node of the tree, and $T$ encodes the reconstructed phylogenetic tree through its branches $(i,j)$.

We assume every position $x$ of the sequence $s$ evolves independently inside each branch. Mutations occur according to a set of Poisson clocks with sequence- and position-dependent rates, $\mu_{s,x}$, expressed per unit time of branch length. During $t$ some positions will mutate and others will remain unchanged, so that
\begin{equation}
\label{model}
P(s \rightarrow s' | t) = \prod_{x\,|\,s(x) \neq s'(x)} \!\!\!(1-e^{-\mu_{s,x}t})  \prod_{x\,|\,s(x) = s'(x)} \!\!\!e^{-\mu_{s,x}t}.
\end{equation}
We assume that
mutability depends independently on the local 5-mer sequence context centered around the mutation, $w(s,x) = (s(x-2),\dots,s(x+2))$, and on the absolute position $x$ along the gene (measured as the distance from the $5'$ end of the gene), so that $\mu_{s,x} = \gamma_{w(s,x)}\beta_x$. In absence of context and position dependence, we would have $\mu=1$ by construction.
Thus values of $\gamma_w$ or $\beta_x$ above 1 imply higher mutabilities than average, and vice versa for values below 1. To lift the degeneracy in overall scale between $\beta_x$ and $\gamma_w$, we impose $\langle\beta_x\rangle=1$.

Overall, the model has $4^5=1024$ parameters for $\gamma_w$ corresponding to each 5-mer, and $L=400$ parameters for $\beta_x$ corresponding to each possible position.
We infer these parameters from repertoire-wide sequencing data by maximizing the total log-likelihood of mutations in all branches in all lineages, $\mathcal{L}(\beta,\gamma)=\sum_{(S,T)} \ln P(S|T,\beta,\gamma)$, with respect to $(\beta,\gamma)$, using an iterative procedure.


\subsection*{Validation on synthetic data}
We first tested the ability of the inference framework to recover true mutability parameters using synthetic datasets. Synthetic data was designed to mimic as closely as possible the features of the real repertoire data to be analyzed.
We used tree topologies inferred on out-of-frame lineages from 9 individuals of Ref.~\cite{Briney2019}. The sequence at the root of each tree was replaced by a random sequence drawn using IGoR, a model of stochastic VDJ recombination \cite{Marcou2018}.
Random mutations were then introduced along the tree structure, following the same number of mutations on each branch as in the original lineage, and according to the SHM model (Eq.~\ref{model}). Context-dependent parameters $\gamma_w$ were set to the previously published S5F model \cite{Yaari13} , and three variants of the position dependent $\beta_x$ were tested: flat, and two sinusoidal profiles (see Methods). Finally we collected sequences at the leaves of the trees into a synthetic dataset.

\begin{figure*}
\begin{center}
  \includegraphics[width=1.\textwidth]{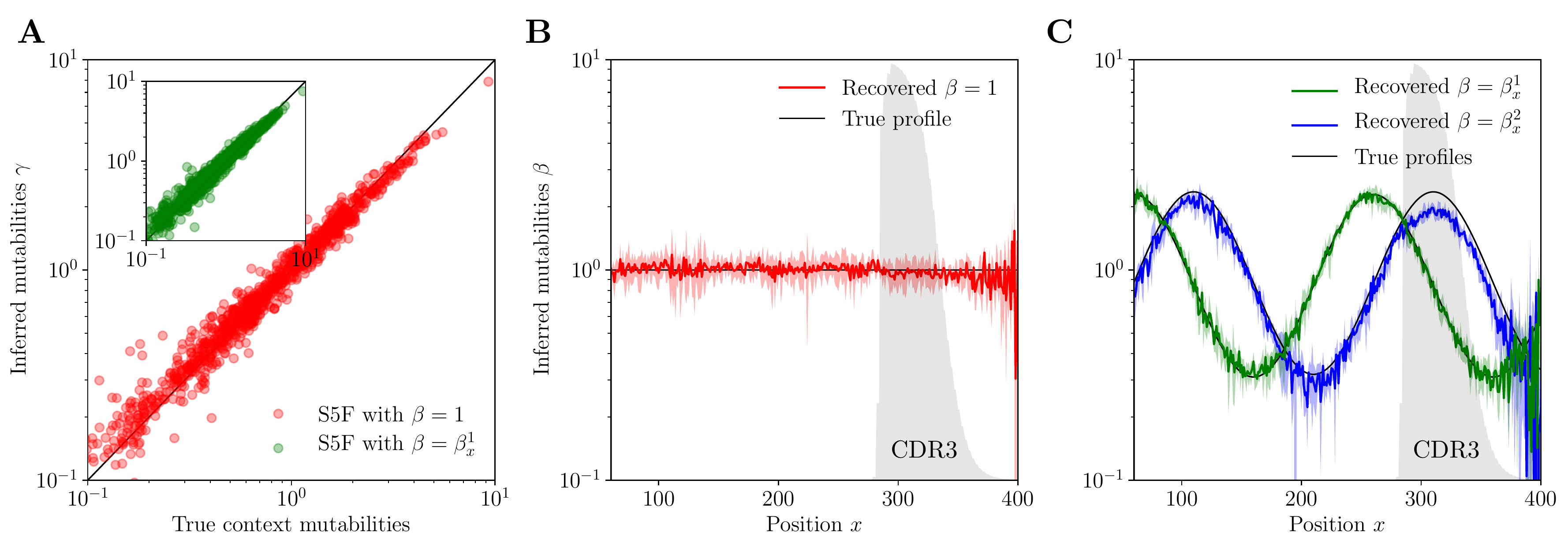}
\end{center}
\caption{Validation of the SHM model inference framework on synthetic data generated with the S5F model \cite{Yaari13}. A. Inference of context mutabilities $\gamma_w$. B,C: Inference of position mutabilities $\beta_x$ for flat (B) and sinusoidal (C) profiles. Error bars correspond to $95\%$ confidence intervals. The frequency at which a given position belongs to the CDR3 region is indicated with the grey shaded area.}
\label{fig:validation}
\end{figure*}

Starting from this dataset, we performed alignment, clonal family inference, tree reconstruction and finally model inference using the exact same procedure as for real data. We compared parameters inferred this way to the true values of $\gamma_w$ and $\beta_x$ (Fig.~\ref{fig:validation}). We were able to recover these rates with excellent accuracy (Pearson's $r^2=97\%$ for both $\gamma$ and $\beta$).

The fact that the procedure recovers the correct position-dependent profile $\beta_x$, including a flat one (Fig.~\ref{fig:validation}B), shows that the framework successfully corrects for the two following confounding factors. First, sequence conservation across the different V, D, and J segments means that context and position are often intertwined, making the extraction of each dependence difficult. Second, high variability in the CDR3 may cause errors in the assignment of sequences into clonal families, and makes it harder to reliably call mutations than in the germline regions. 
This remains true in the presence of large variations of the mutability along the position, including in the CDR3, as demonstrated on the sinusoidal profiles (Fig.~\ref{fig:validation}C).
On the other hand, the possibility to use the CDR3 sequence for model inference gives access to a more diverse range of possible contexts, leading to better estimates for contexts that are underrepresented in the germline genes.

To assess the impact of errors in the reconstruction of clonal families and lineages on the inferred parameters, we repeated the procedure using the true tree topologies instead of the reconstructed ones. This only modestly improved accuracy ($r^2=98\%$, see Fig.~S1), suggesting that the procedure is robust to lineage misassignments.

\subsection*{Mutabilities depend on both sequence context and position}
\begin{figure*}
\begin{center}
\includegraphics[width=1.\textwidth]{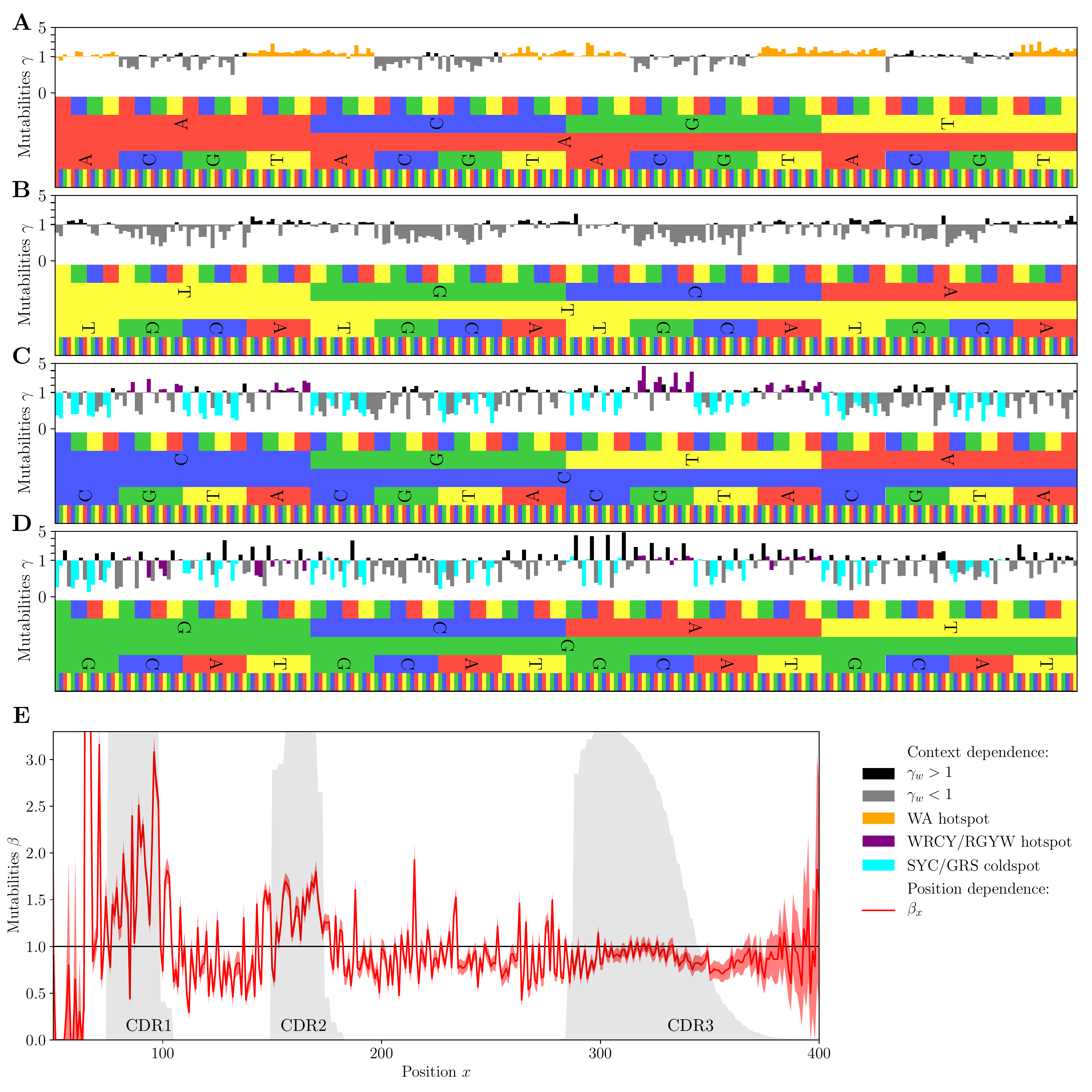}
\end{center}
\caption{Context and position dependent model parameters. Context-dependent mutabilities $\gamma_w$ for A- (A),T- (B),C- (C) and G-centered (D) 5-mers. The colors indicate known hopspot and coldspot motifs.
  E. Position-dependent mutabilities $\beta_x$. Gray shadings show the probability to be in the CDR1, CDR2, or CDR3 regions. Error bars correspond to $95\%$ confidence intervals. See Fig.~S5 for an full analysis of parameter uncertainty.}
\label{fig:results}
\end{figure*}

Confident that our procedure is able to infer rates reliably, we next applied it to real data, consisting of the out-of-frame lineages from Ref.~\cite{Briney2019}. The inferred dependencies of mutability with context and position are presented in Fig.~\ref{fig:results}. We represent context dependence using a flat variant of the ``hedgehog'' plots used in Ref.~\cite{Yaari13}, for A-, T- , C-, and G-centered motifs (Fig.~\ref{fig:results}A-D). Full parameter tables are available at \url{https://github.com/statbiophys/shmoof}.

Context dependent rates for A-centered motifs correspond well to the standard W\underline{A} classification \cite{Zhao8146}: $76\%$ of A-centered 5-mers with $\gamma_w>1$ are of the W\underline{A}  type, and only 7 of 128 W\underline{A} 5-mers have $\gamma_w\lesssim 1$. T-centered motifs are dominated by coldspots and their mutabilities do not align well with their corresponding reverse complement counterparts. This is in agreement with the known property of Polymerase $\eta$ to be prone to errors at A nucleotides on the top strand \cite{Pilzecker2019}.

The C- and G-centered motifs have largely reverse-complement-symmetric rates (see Fig.~S2). As previously noted \cite{Yaari13}, this is in agreement with the strand-symmetric targeting of C/G-centered motifs by the AID enzyme.

The previously reported WR\underline{C}Y/R\underline{G}YW motif \cite{Unniraman2007,Feng2020} predicts high mutability reasonably well, while the SY\underline{C}/\underline{G}RW class of motifs \cite{pham2003} explains well a good fraction of coldspot motifs. Importantly, a large number of high or low mutability 5-mers do not belong to any of the previously reported motifs (see Supplementary Tables~1 and~2).

The rugged profile of position dependence (Fig.~\ref{fig:results}E) shows clear enrichment in mutations in the CDR1 and CDR2 regions, reflected in the up to 2-fold increase of the position-dependent rates. Framework regions are less mutated and we also observe a slight drop in the mutabilities of the positions beyond the Cysteine anchor of the CDR3 region. We also learned models where the position was defined from the 3' end of the sequence in the J segment (Fig.~S3), yielding similar results but no clear improvement over 5'-based position. High mutability of CDR1 and CDR2 was already noted \cite{Saini2015} and justified as an enrichment in highly mutable motifs (as quantified with the S5F model). Our findings suggest that there is a secondary mechanism of this enrichment, having to do either with accessibility of mutation-inducing enzymes or a superposition of context-dependent effects that evade the assumption of independent evolution at different sites and the limitation of 5-mer motifs. 

Note that introducing the explicit position dependence does affect the learning of the context-dependent parameters: learning $\gamma_w$ with no position dependence (fixing $\beta_x=1$) yields similar but markedly different parameters than when learning a free $\beta_x$ ($r^2=81\%$, Fig.~S4).

\begin{figure*}
\begin{center}
\includegraphics[width=.9\textwidth]{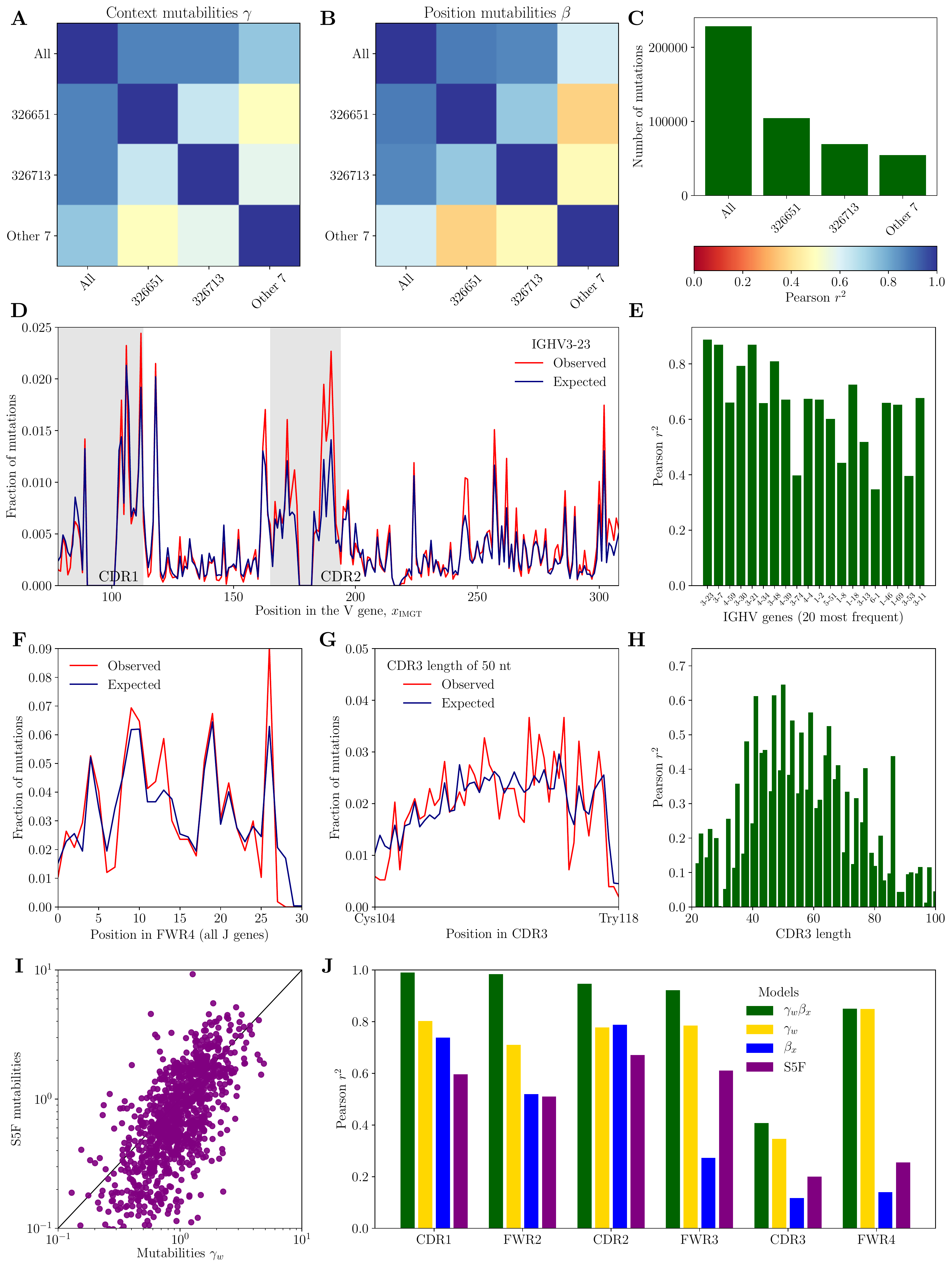}
\end{center}
\caption{
The model explains the data. Observed profiles were measured across the entire dataset used for model inference. See Fig.~S6 for an equivalent figure when data was divided into training and testing sets.  
Reproducibility of parameters for individual-specific models: context (A) and position (B) mutabilities. 
C. Number of mutations used for inference 
D. An example mutation profile in the most common V gene. 
E. Model performance across V genes.
F. Mutation profile in the FWR4 region.
G. An example mutation profile in the CDR3 region for CDR3 length of 50 nts.
H. Model performance across CDR3 lengths.
I. Comparison with the S5F model.
J. Summary of models performance across sequence regions. We compare the full context- and position-dependent model ($\gamma_w\beta_x$) with purely context- ($\gamma_w$) and position- ($\beta_x$) dependent models as well as with the S5F default model.
}
\label{fig:explain_data}
\end{figure*}

\subsection*{Model is consistent across individuals and explains data better than previous approaches} 
To check the model's generality, we estimated its variability across individuals by computing Pearson's correlation coefficient between the context ($\gamma_w$, Fig.~\ref{fig:explain_data}A) and position ($\beta_x$, Fig.~\ref{fig:explain_data}B) mutability profiles of different donors. The precision with which we can estimate model parameters depends on the number of sequences used for inference, particularly for rare 5-mer contexts. 
Because two individuals had many more reads than the others 7, we pooled together these seven individuals to make comparisons with similar dataset sizes (Fig.~\ref{fig:explain_data}C). We then compared the 2 individuals and 1 meta-individual with each other and with a model learned on data from all individuals.
For the 2 individuals with the largest repertoire datasets, the results are highly reproducible with Pearson's $r^2=78\%$ for context and $r^2=70\%$ for position parameters (Fig.~\ref{fig:explain_data}A), suggesting that the model captures universal biochemical properties of the hypermutation process.

To further validate the model's accuracy, we compared its prediction to data on the V-specific mutation profiles, which consist of the position-dependent mutation rate for each V segment. These rates result from the combined effect of position and context, but they are not fitted directly by the model. A typically good example of such a profile is shown in Fig.~\ref{fig:explain_data}D. The prediction is generally excellent (Pearson's $r^2\sim 50-80\%$), and is poorest for V segments for which little data was available (Fig.~\ref{fig:explain_data}E). Similarly, the model predicts well the mutability on Framework Region 4 (FWR4), which encompasses the J segment (Fig.~\ref{fig:explain_data}F), as well as in the CDR3 (Fig.~\ref{fig:explain_data}G and H), which is usually ignored in other approaches. Performance is best for the most frequent CDR3 length (Fig.~\ref{fig:explain_data}H).

We compared the results of our inference to the S5F model \cite{Yaari13}, which was trained on independent data. The S5F model is defined by a mutability table $\gamma_w$ with no attempt to disentangle position dependence, so a direct comparison is subject to caution. Besides, S5F mutabilities are learned from synonymous mutations of productive sequences, requiring extrapolation methods to cover all $1024$ contexts, all of which do not occur with synonymous mutations. Yet, the two sets of mutabilities $\gamma_w$ correlate fairly well ($r^2=36\%$, Fig.~\ref{fig:explain_data}I). Correlation rises to $r^2=44\%$ for contexts appearing in synonymous mutations, versus $r^2=18\%$ for the other contexts for which S5F recourses to extrapolation, emphasizing the limitations of that extrapolation.

A summary of the performances of the different modeling approaches on the mutabilities in the different regions of the IgH gene is shown in Fig.~\ref{fig:explain_data}J. We also checked for overfitting by dividing the dataset into a training and a testing, finding similar results (Fig.~S6). 
The full position and context dependent model ($\mu_{s,x}=\gamma_w\beta_x$) performs better than models with context or position alone ($\mu_{s,x}=\gamma_w$ and $\mu_{s,x}=\beta_x$). While the context explains the bulk of the mutation profile, adding positional effects substantially boosts performance. Our model clearly outperforms the S5F model, although it should be reminded that S5F was trained on a distinct dataset. Re-training S5F on the productive sequences from the present datasets using the procedure described in the original article \cite{Yaari13} actually yielded worse performance (data not shown), for reasons that are unclear to us. Overall, accounting for phylogeny and disentangling the combined effects of context and position allows our model to accurately predict mutabilities including in the hyper-variable CDR3 region.

\subsection*{Co-localization of mutations cannot be explained by context and position bias}
It was previously observed that hypermutations tend to cluster along genomic position in nonproductive sequences \cite{Marcou2018}. However, the origin of this phenomenon and its dependence on confounding factors such as phylogeny and heterogeneous hot spot concentration were not fully characterized.
 
\begin{figure*}
\begin{center}
  \includegraphics[width=\textwidth]{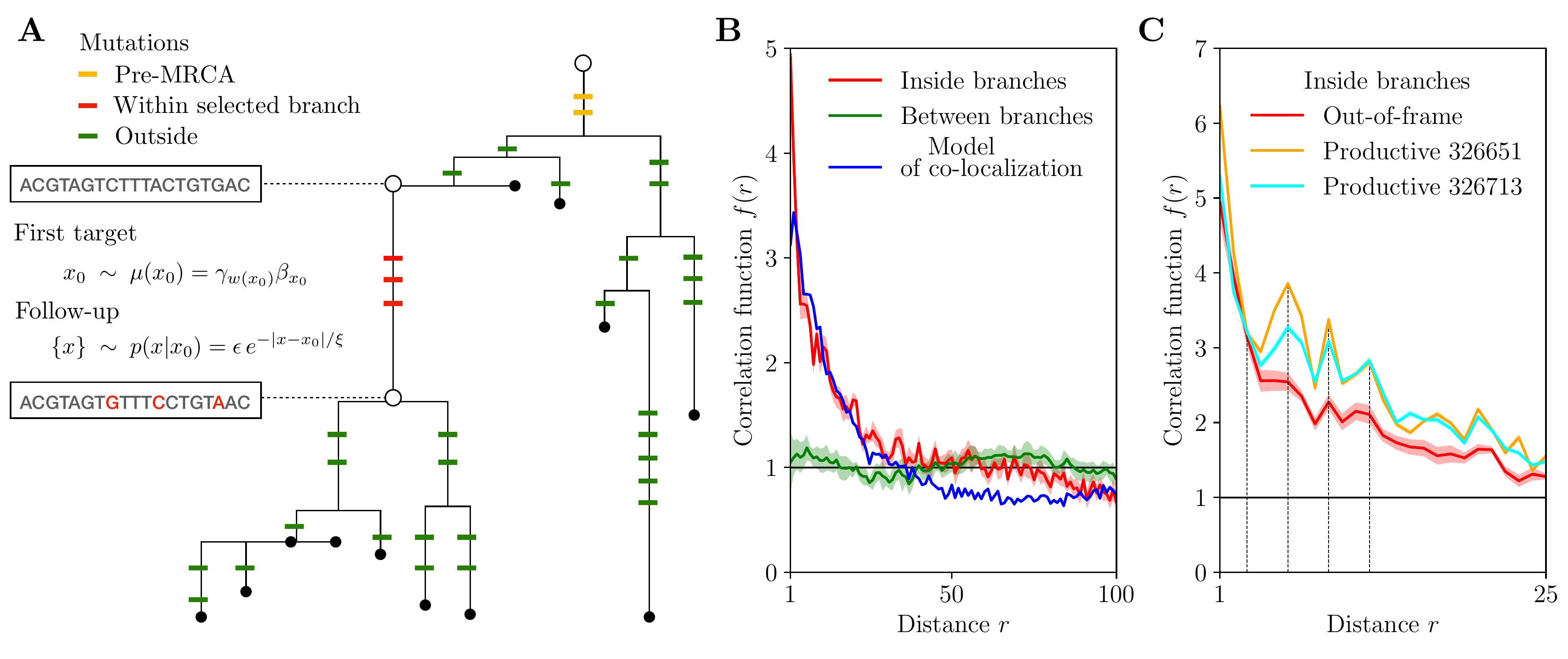}
\end{center}
\caption{Co-localization of subsequent hypermutations. A. Co-localization model explained on an example phylogeny. An initial mutation is drawn from the context- and position-dependent model. Then, follow-up mutations are drawn in its vicinity within the same branch.  B. Correlation function $f(r)$ for pairs of mutations: inside (red) and between (green) branches and  for simulated mutations according to the co-localization model (\ref{new_toy}) with $\epsilon=5\%$ and $\xi=10$. C. Co-localization of subsequent hypermutations in productive lineages from the 2 largest individuals' datasets (ID=326651 and ID=326713). Correlation function $f(r)$ for pairs of mutations inside branches compared to the out-of-frame result. The first multiplicities of the codon frame length, $r=3,6,9,12$ are marked with dotted lines to guide the eye. Shaded areas represent $95\%$ confidence intervals.}
\label{fig:correlation}
\end{figure*}

Clustering of mutations can be directly observed by plotting the fraction $n(r)$ of pairs of mutations at distance $r$ from each other as a function $r$ (normalized by the total number of pairs at that distance, see Fig.~S7), which is also called a spatial correlation function in physics. Focusing on lineages with at least 6 leaves, and iterating through all branches with fewer than 10 mutations, we evaluated this correlation function for pairs of mutations occurring in the same branch of the phylogeny {\em versus} distant branches, as schematized in Fig.~\ref{fig:correlation}A).
We then compared this correlation function to our model predictions (Fig.~S8). 
{\revision
The enrichment of closeby mutations can be quantified by the correlation function $f(r)={n(r)}/{n_m(r)}$, where $n(r)$, the fraction of pairs of mutations distant by $r$ in the same tree branch is normalized by the model prediction $n_m(r)$ (Fig.~\ref{fig:correlation}B).
}

Pairs of mutations in distinct branches are well explained by the model, suggesting that they are independent of each other, in agreement with the biological picture that they occur at different rounds of affinity maturation. The enrichment of closeby mutations in distant branches can be entirely explained by the clustering of hotspot regions. Interestingly, both context and position dependencies of the mutability are needed to explain the data (Fig.~S8).
In contrast, pairs of mutations inside branches tend to occur closer to each other than predicted by the model.
The enrichment of closeby mutations is up to five-fold, pointing to an additional mechanism causing hypermutation clustering. 
We observe that this enrichment persists in the presence of selection, as verified by computing the correlation function $f(r)$ in productive lineages (Fig.~\ref{fig:correlation}C).

\subsection*{Minimal model of co-localization}
To explain the observed excess of co-localized mutations, we propose a simple phenomenological model. Targeted mutations, following the context and position dependent profiles described so far, cause additional nearby `follow-up' mutations due to error-prone DNA repair. 
Given a substitution at $x_0$ drawn from the same distribution as before, each position $x\neq x_0$ can subsequently mutate with probability 
\begin{equation}
\label{new_toy}
    p(x|x_0) = \epsilon \, e^{-|x-x_0|/\xi},
\end{equation}
where $\xi$ is the correlation length and $\epsilon$ is small. The total number of follow-up mutations is approximately Poisson distributed with mean 
$\sum_{x} p(x|x_0)\simeq {2\epsilon}/{(1-e^{-1/\xi})}$.
To simulate this process, we followed the same procedure as described earlier for synthetic data, but allowing for follow-up, as well as targeted mutations, while keeping the total number of mutations in each branch constant.
We then computed the correlation function $f(r)$, and compared it to true profiles (Fig.~\ref{fig:correlation}B). We obtain a good agreement for $\xi=10$ and $\epsilon=5\%$ corresponding to an average of $\sim 1$ follow-up mutation per targeting event. This result suggests that as many as $50\%$ of observed mutations are follow-up mutations.

{\revision
  We asked whether this large number of non-targetted mutations may bias the inference of the targeting model, which assumes no follow-up mutations. To assess this effect, we re-inferred the rates $\{\beta_x\}$ and $\{\gamma_w\}$ from synthetic datasets simulated with $\xi=10$, $\epsilon=5\%$, with data-inferred context profile $\gamma_w$, and with data-inferred or flat position profiles $\beta_x$ (Fig.~S9). We find that the re-inferred mutabilities mostly agree with the true ones, with a slight shrinkage of values and enhanced mutabilities of cold spots, owing to the equalizing effect co-localization.
  Importantly, co-localization does not introduce additional features in the re-inferred position-dependent profile, indicating that our inference procedure is robust to co-localization effects.
}

\section{Discussion}

The mutational landscape of antibody repertoires results from many entangled effects, which are often lumped together into effective models of hypermutations \cite{Yaari13,Schramm2018,Dhar2019}.
First, hypermutations have intrinsic preferences for certain positions along the IgH gene, regardless of their impact on protein function. In addition, selection for antibody function, which includes protein stability and antigen affinity, favors beneficial mutations and suppresses deleterious ones \cite{Feng2020}. While intrinsic SHM preferences are believed to be universal, selective forces vary across lineages which are involved in distinct immune responses \cite{Dhar2018}, and may also depend on the individual's immune status \cite{Zuckerman2010}.
Repertoire sequencing gives a snapshot of a rapidly adapting population subjected to these forces, making it hard to disentangle intrinsic SHM preferences from the combined effects of selection and genetic drift. By focusing on non-productive lineages and using a phylogeny-based approach, we overcome the biases arising from the dynamics of affinity maturation to obtain a comprehensive picture of SHM intrinsic preferences.

Each hypermutation occurs through a series of events of DNA damage and repair. The action of each enzyme, including AID to error-prone DNA repair enzymes, may each have their own sequence preferences, and the interplay of these different biases results in the observed profile.
In our approach, these complex mutational pathways are subsumed into an effective model with a limited number of interpretable parameters in terms of effective  context and position dependence. As a result, the context dependent weights $\gamma_w$ do not simply reflect the binding preference of AID, but also account for the biases of other biochemical steps. Our framework enables direct measurement of the mutability $\gamma_w$ of a wide range of 5-mer contexts, recovering the known classifications of hot and cold spots \cite{Unniraman2007,Yaari13}. We show that our model outperforms existing methods as well as purely context or position dependent models in terms of explaining the data.

The introduction of an explicit and universal position dependence, $\beta_x$, allows us to unveil an excess of mutations in the CDR1 and CDR2 regions. This enrichment of mutations cannot be simply explained by their harboring more hotspot contexts. We cannot exclude that this residual position dependence is due to  more complex context effects missed by our model (based e.g. on 7-mers, which would be impractical to infer from the present dataset). 
Alternatively, SHM may preferentially target these regions independently of their sequence context, possibly through epigenetic mechanisms. Such preference is known to exist at the genome-wide level to mutate the Ig loci without affecting other genes \cite{Feng2020}, so it is plausible that the same mechanism targets some specific positions within Ig. The enrichment of mutations in the CDR1 and CDR2 regions is even more marked in productive sequences, meaning that these mutations are more likely to be selected during affinity maturation. This suggests that the intrinsically enhanced mutability of these regions may carry an evolutionary advantage, by focusing hypermutations on regions where they are the most beneficial \cite{Saini2015}.
The stability of the immunoglobulin relies on the FWR regions and most of the substitutions are expected to be deleterious. The purifying nature of selection in FWR regions has been quantified in Ref.~\cite{Nourmohammad2019} and contrasted with positive selection in CDR regions.

By studying mutations along lineages, we were able to study mutations in the probable context in which they occurred, rather than relative to the germline sequence, allowing us to take into account the order of mutations and to sample a broader diversity of 5-mer contexts. This approach also allowed us to study and characterize hypermutations in the CDR3, which has been neglected in previous work \cite{Bonsignori2016} owing to the difficulty to separate these mutations from junctional diversity.

The phylogenetic methods employed in this study were not specifically designed to study B cell repertoires. In particular the assumptions allowing for fast likelihood computations do not account for the context dependence of the mutation rate beyond the codon frame \cite{Hoehn2019}. The position-dependent model introduced here could offer a compromise. While it does not account for the the full complexity of SHM biases, it captures the variation of the mutation rate observed in out-of-frame data well (Fig.~\ref{fig:explain_data}), and can operate under the assumption of independent site evolution. Our framework could also be easily extended to include position-dependent selection in the nucleotide or amino acid representation. 

Our analysis confirmed a phenomenon of co-localization of mutations along the sequence. While this effect had been previously reported \cite{Marcou2018}, here we showed that it could not be explained by phylogenetic bias or the existence of regions of higher and lower mutabilities. We proposed a minimal quantitative model of hypermutation targeting, followed by error-prone DNA repair that causes follow-up mutations, which explains the data well. While ideally we would like to infer the position and context mutability profiles taking these follow-up mutations into account, the task is impractical because it would require to identify the origin of each mutation. We expect that doing so would only renormalize the values of the context preferences. 
While the adaptive advantage of co-localized mutations is unclear, we find the correlation function in productive lineages follows the unproductive baseline with additional enrichment  enhanced at multiples of the codon length, 3, suggesting signatures of selection (Fig.~\ref{fig:correlation}C). 
We speculate that nearby mutations occurring simultaneously could help cross barriers of positive sign epistasis, whereby two or more mutations are deleterious by themselves, but beneficial together. This phenomenon could accelerate affinity maturation by favoring compensatory or epistatic mutations at amino acids that interact strongly within the antibody protein \cite{Koenig2015,Adams2019}.

The obtained mutability models make predictions about the likelihood and plausibility of particular trajectories of affinity maturation. They could be useful in designing vaccination strategy, by helping choose targets with a greater potential for accumulating beneficial mutations towards antibodies with desired properties such as neutralization power, or broadness in the case of fast evolving pathogens such as influenza or HIV \cite{Liao2013,Bonsignori2016}.

\section{Methods}

\subsection*{Data and preprocessing}
We perform the analysis on recently published high-throughput RNA sequencing of Ig heavy genes at great depth \cite{Briney2019}. 

The sequences were barcoded with unique molecular identifiers (UMI) to correct for the PCR amplification bias. However, UMI cannot be used to correct sequencing errors, as most UMI were represented by a single sequence: the number of UMIs used is of the same order as the total number of cells in use. We aligned raw sequences using presto of the Immcantation pipeline \cite{VanderHeiden2014} with setup allowing to correct for errors in UMIs and deal with insufficient UMI diversity. The V region primers were masked and the C region primers were used to distinguish the two isotypes of sampled B cells: the IgM and IgG classes. The study of mutation profiles in the two groups revealed a much lower mutational load in the IgM cohort and hence a higher relative level of sequencing errors, as well as shallower tree topologies. For further analysis we chose to focus exclusively on the IgG class. Reads were filtered for quality and paired using default presto parameters. Pre-processed data was then aligned to V, D and J templates from IMGT \cite{IMGT} database using IgBlast \cite{Ye2013}.
In total there were $3.6\,\times 10^{6}$ IgG sequences per person (average $3.6\times 10^{6}$, median $1.8\times 10^{6}$), of which up to $2\%$ were unproductive (average $5.7\times10^4$, median $2.9\times10^4$).

\subsection*{Inference of evolutionary trajectories}
Sequences with a frameshift in the CDR3 region were then selected and used to reconstruct clonal families as follows. 
In the first step, reads were aligned to the V and J templates and grouped into classes of sequences with the same V and J gene assigned, as well as equal CDR3 length. In the out-of-frame classes we inferred clonal lineages by single linkage clustering with a threshold of $90\%$ on CDR3 region identity \cite{Gupta2015}. We reconstruct maximum likelihood topologies, as well as the identity of ancestral states, under a simple K80 model of character evolution \cite{Kimura1980} for all lineages comprising at least 6 unique sequences. The model does not capture the complexity of the observed mutation profile, but avoids fitting multiple parameters independently in small lineages of relatively short alignment. The existing repertoire-wide method \cite{Hoehn2019} is incompatible with out-of-frame lineages since it operates on 61 productive codons. Ancestral states are found through marginal reconstruction. Germline V and J sequences were used as an outgroup to inform the phylogenetic inference and root the lineage.

\subsection*{Model inference}
With the exception of the initial branch, which joins the germline sequence and the most recent common ancestor of the lineage, all branches shorter than 10 substitutions were used for model inference.

Our task is to find a set of parameters $\{\gamma_w\},\{\beta_x\}$ that maximise the log-likelihood
\begin{equation}
  \label{total_likelihood2}
  \begin{split}
    \mathcal{L} =
    \sum_{S,T;(i,j)\in T} &\left[\sum_{x'\,|\,s_i(x')\neq s_j(x')}\ln\left(e^{\gamma_{w(s_i,x')}\beta_{x'} t} - 1\right)\right.\\
    &\left.- \sum_{x} \gamma_{w(s_i,x)}\beta_x t\right],
    \end{split}
\end{equation}
where $S$ is the set of sequences (observed and reconstructed) at each node of the tree, and $T$ encodes the reconstructed phylogenetic tree through its branches $(i,j)$, with reconstructed ancestral states $s_i$ and $s_j$. The rates $\mu_{s,x}=\beta_x\gamma_{w(s,x)}$ are defined so that the length of each branch $t$ is expressed in terms of the expected number of substitutions per basepair (total number of substitutions divided by the total alignment length).
Imposing $\partial \mathcal{L}/\partial \gamma_w=0$ yields an implicit expression for $\gamma_w$ as a function of $\{\beta_x\}$, but independent of $\{\gamma_{w'}\}_{w'\neq w}$, which can be solved by one-dimensional root finding. Likewise, setting $\partial \mathcal{L}/\partial \beta_x=0$ gives an implicit expression for $\beta_x$ as a function of $\{\gamma_w\}$.
We can perform the following iteration:
\begin{align}
\label{maximize}
    \gamma^{n} &=  \underset{\gamma}{\arg\max}\;\mathcal{L}(\gamma,\beta^{n}) \\
    \beta^{n+1} &= \underset{\beta}{\arg\max}\; \mathcal{L}(\gamma^n,\beta),
\end{align}
which converges to the maximum of $\mathcal{L}$ with respect to the joint $\{\gamma_w,\beta_x\}$.

To estimate the uncertainty of inferred parameters we sample with replacement from the set of all branches to create 400 bootstrap copies. We report $95\%$ confidence intervals.

\subsection*{Substitution models}
Not only the targeting rate, but also the identity of the substitution is known to depend on the identity of neighboring bases \cite{Yaari13}. In our formulation of the model, inference of the targeting rates does not require knowing the substitution type, however we can easily extend the framework to include this dependence. The probability of mutating from $w$ to $w'$ over a period $t$ can be expressed as
\begin{equation}
\label{subst_model}
P(w \rightarrow w' | t) = 1-e^{-\gamma_{w} \omega_{w\rightarrow w'} \beta_{x} t},
\end{equation}
where $\sum_{w'} \omega_{w\rightarrow w'} = 1$ and $\omega_{w\rightarrow w'} \neq 0$ if  $w'$ is a result of a substitution at the central position of $w$. This way we add $2\times4^5=2048$ parameters to the model. We can infer the maximum likelihood estimates of $\omega_{w\rightarrow w'}$ using the same iterative scheme introduced in the previous section.

\subsection*{Synthetic datasets}

We created synthetic datasets using the S5F model of mutability (downloaded from \url{clip.med.yale.edu/shm}) for $\gamma_w$. We used a flat profile, $\beta_x=1$ as well as sinusoidal profiles $\ln\beta^1_x=2\sin(x/\delta)-1$ and $\ln\beta^2_x=2\cos(x/\delta)-1$ with $\delta=50$. For each branch $(i,j)$, we compute the mutability $\mu_{s_i,x}$ as a function of $x$, and then introduce mutations at $n$ random positions picked without replacement according to $\mu_x/\sum_{x'} \mu_{x'}$, where $n$ is the number of mutations on the branch (fixed by the lineage structure taken from the real data).

\section*{Data availibility}
All the data analyzed in this paper has been previously published and can be accessed from original publications.
Code for producing the figures of this paper, as well as the inferred model parameters, are freely available at \url{https://github.com/statbiophys/shmoof}.

\section*{Acknowledgements}
The study was supported by the European Research Council COG 724208. The authors are grateful for the discussions and suggestions from Thomas Dupic, Quentin Marcou and Victor Chard\`es.

\bibliographystyle{pnas}

\onecolumngrid

\newpage

\section*{Supplementary information}

\renewcommand{\thefigure}{S\arabic{figure}}
\setcounter{figure}{0}

\begin{figure*}[h]
\begin{center}
  \includegraphics[width=\textwidth]{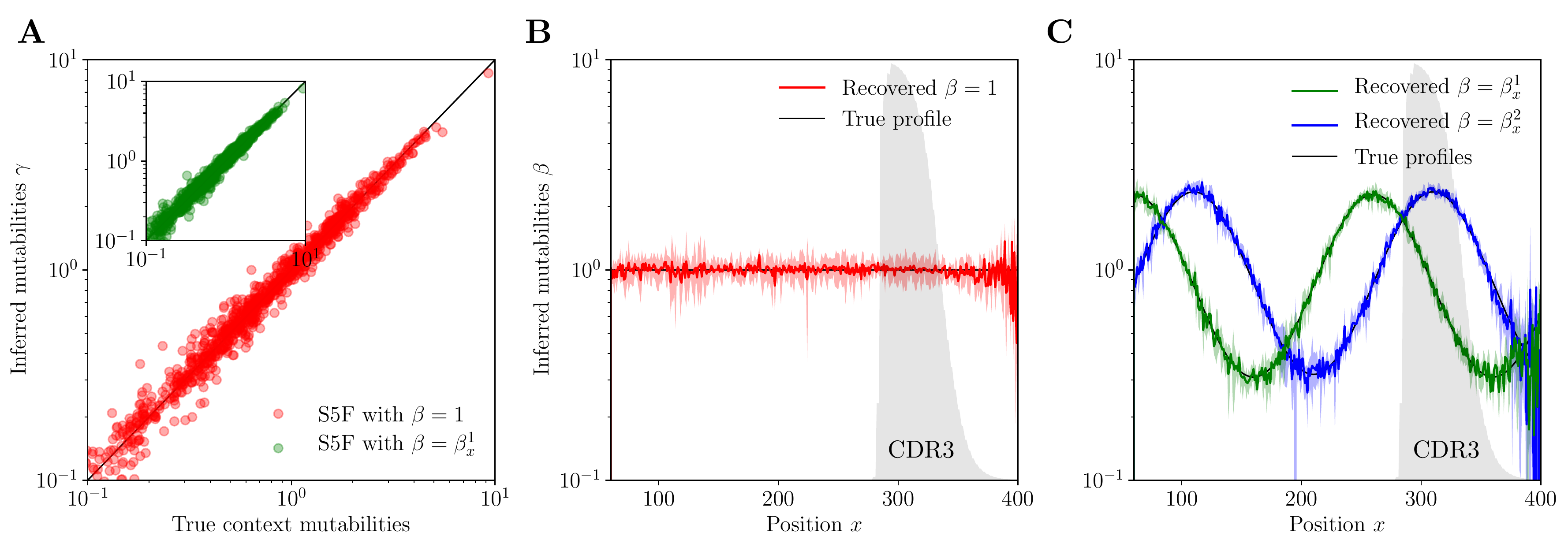}
\end{center}
\caption{Model inference on synthetic data using true phylogenies. A. Inference of context mutabilities $\gamma$. B,C. Inference of position mutabilities $\beta$ for flat and sinusoidal profiles. Error bars correspond to $95\%$ confidence intervals. Frequency at which a given position belongs to the CDR3 region is indicated with the grey shaded areas.}
\label{fig:validation_known_trees}
\end{figure*}

\begin{figure*}
\begin{center}
  \includegraphics[width=\textwidth]{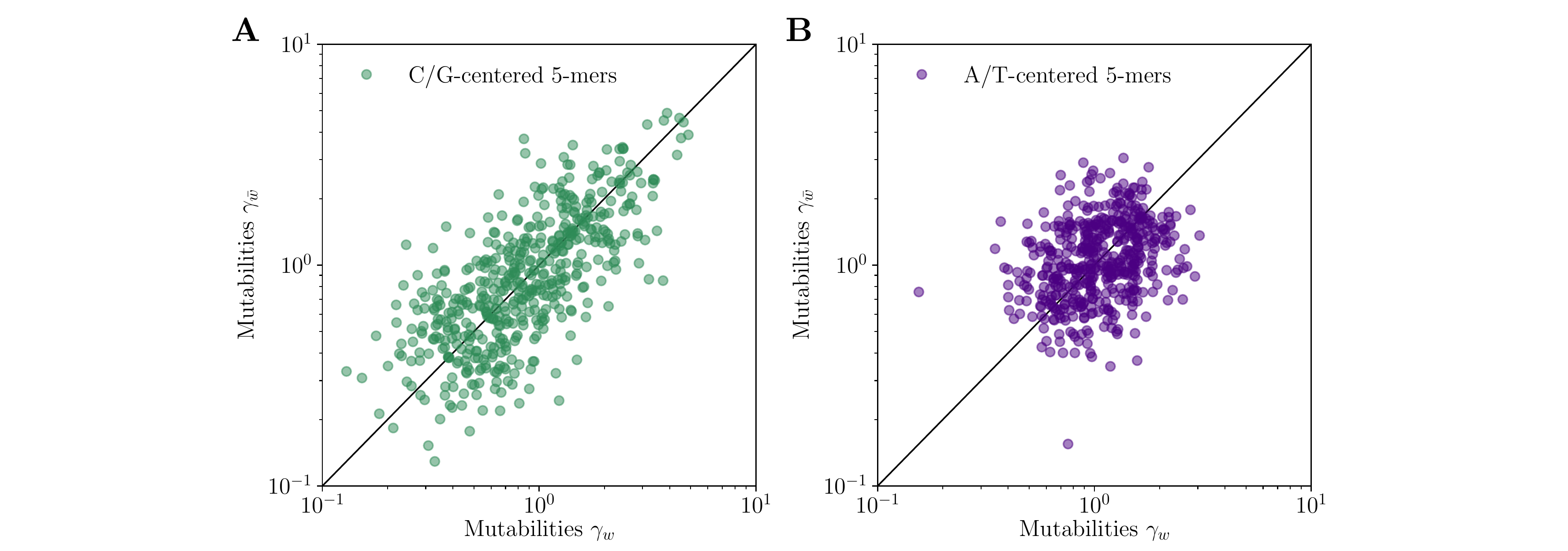}
\end{center}
\caption{Strand asymmetry of the context-dependent rates. For each motif $w$ we juxtapose its mutability $\gamma_{w}$ with the mutability of its reverse complement $\gamma_{\bar{w}}$. A. 5-mer motifs with strong central nucleotide (C/G), $r^2=54\%$. B. 5-mer motifs with weak central nucleotide (A/T), $r^2=7\%$}
\label{fig:strand_assymetry}
\end{figure*}

\begin{figure*}
\begin{center}
  \includegraphics[width=\textwidth]{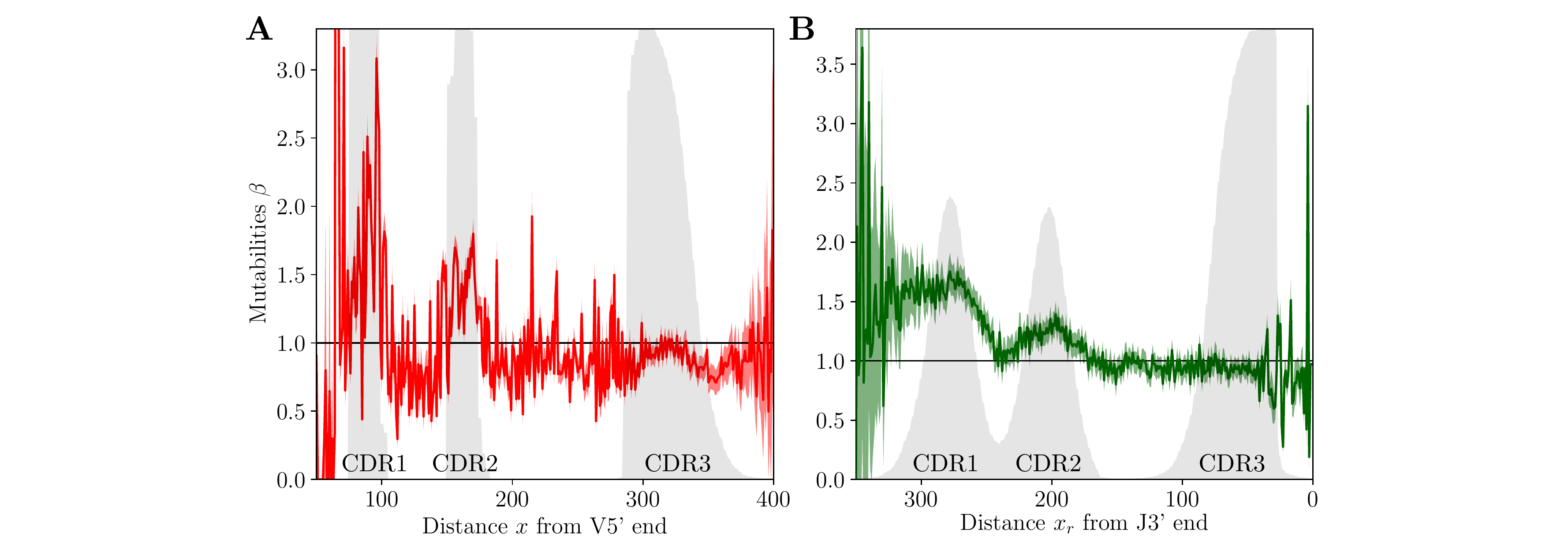}
\end{center}
\caption{Alternative position definition. We compare the models based on different position definitions: $x$, the distance from the $5'$ end of the V gene (A, red) and $x_{r}$, distance from the $3'$ end of the J gene (B, green). Error bars correspond to $95\%$ confidence intervals. Frequency at which a given position belongs to a CDR region is indicated with the grey shaded areas.}
\label{fig:position_right}
\end{figure*}

\begin{figure*}
\begin{center}
  \includegraphics[width=\textwidth]{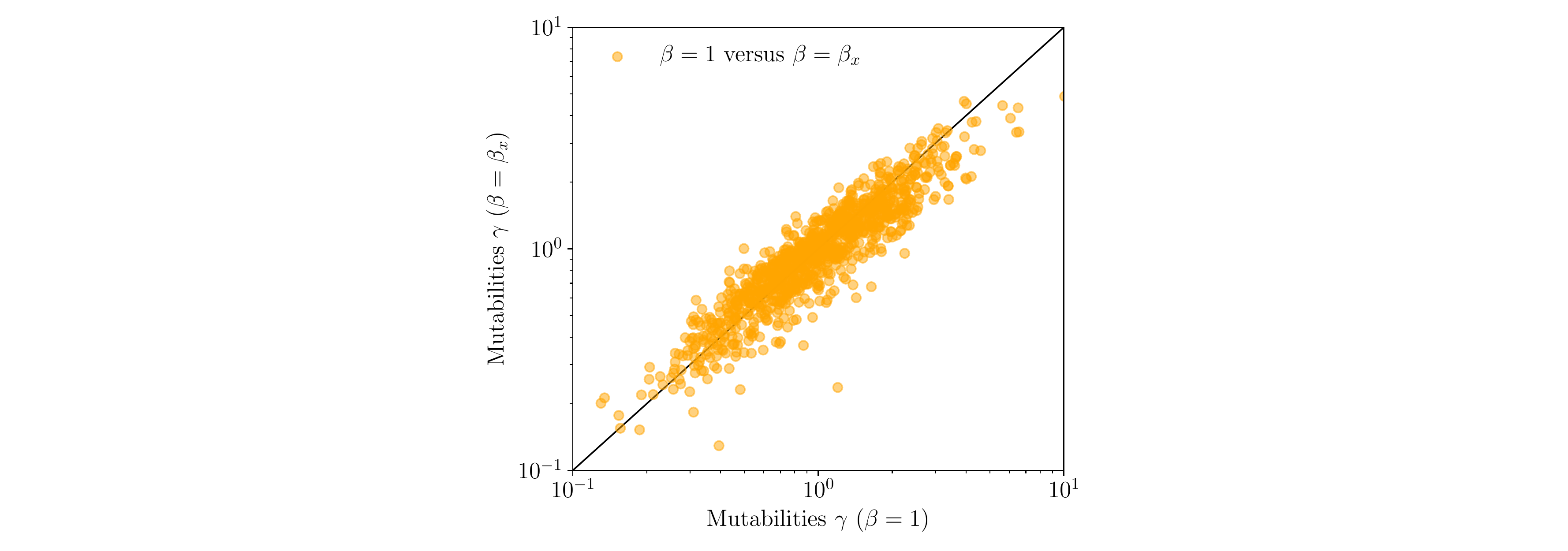}
\end{center}
\caption{Introducing explicit position dependence $\beta \neq 1$ influences the context-dependent rates. We compare the $\gamma$ mutabilities from the full model with the parameters of the purely context-dependent model.}
\label{fig:position_changes_context}
\end{figure*}

\begin{figure*}
\begin{center}
  \includegraphics[width=\textwidth]{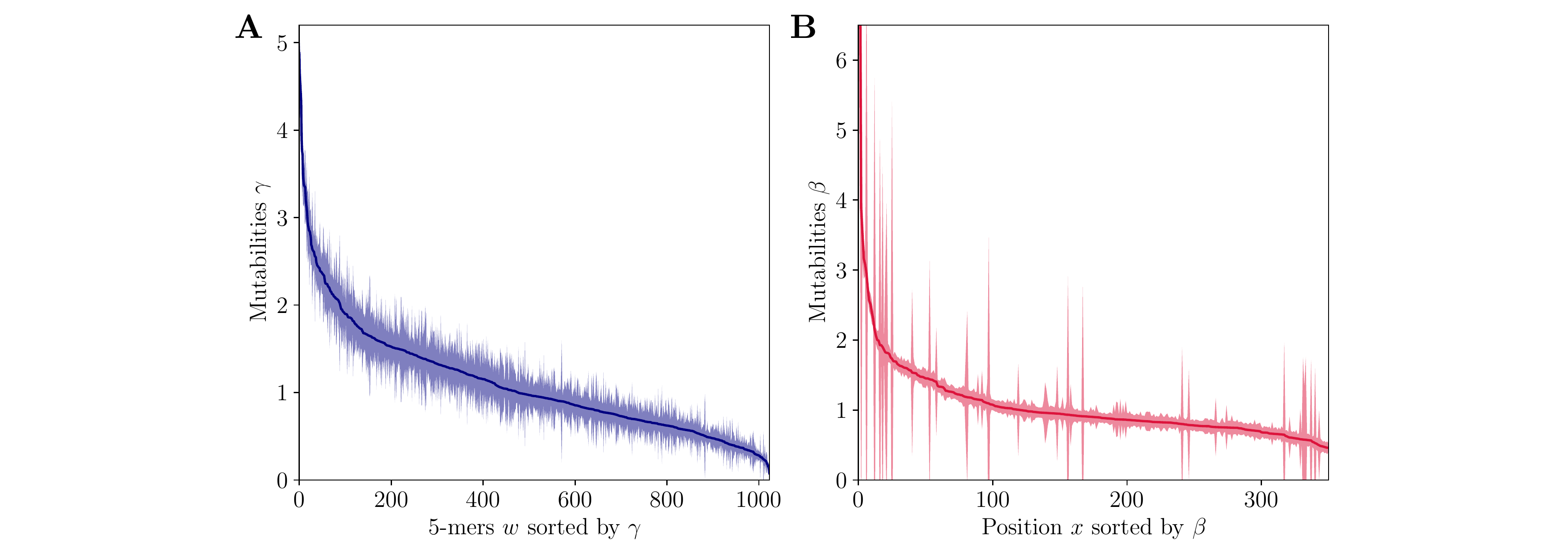}
\end{center}
\caption{Analysis of the uncertainty of estimate parameters $\gamma$ (A) and $\beta$ (B). The shaded area indicates the $95\%$ confidence interval envelope. Motifs and positions are sorted by their respective mutabilities}
\label{fig:parameters_confidence}
\end{figure*}

\begin{figure*}
\begin{center}
  \includegraphics[width=\textwidth]{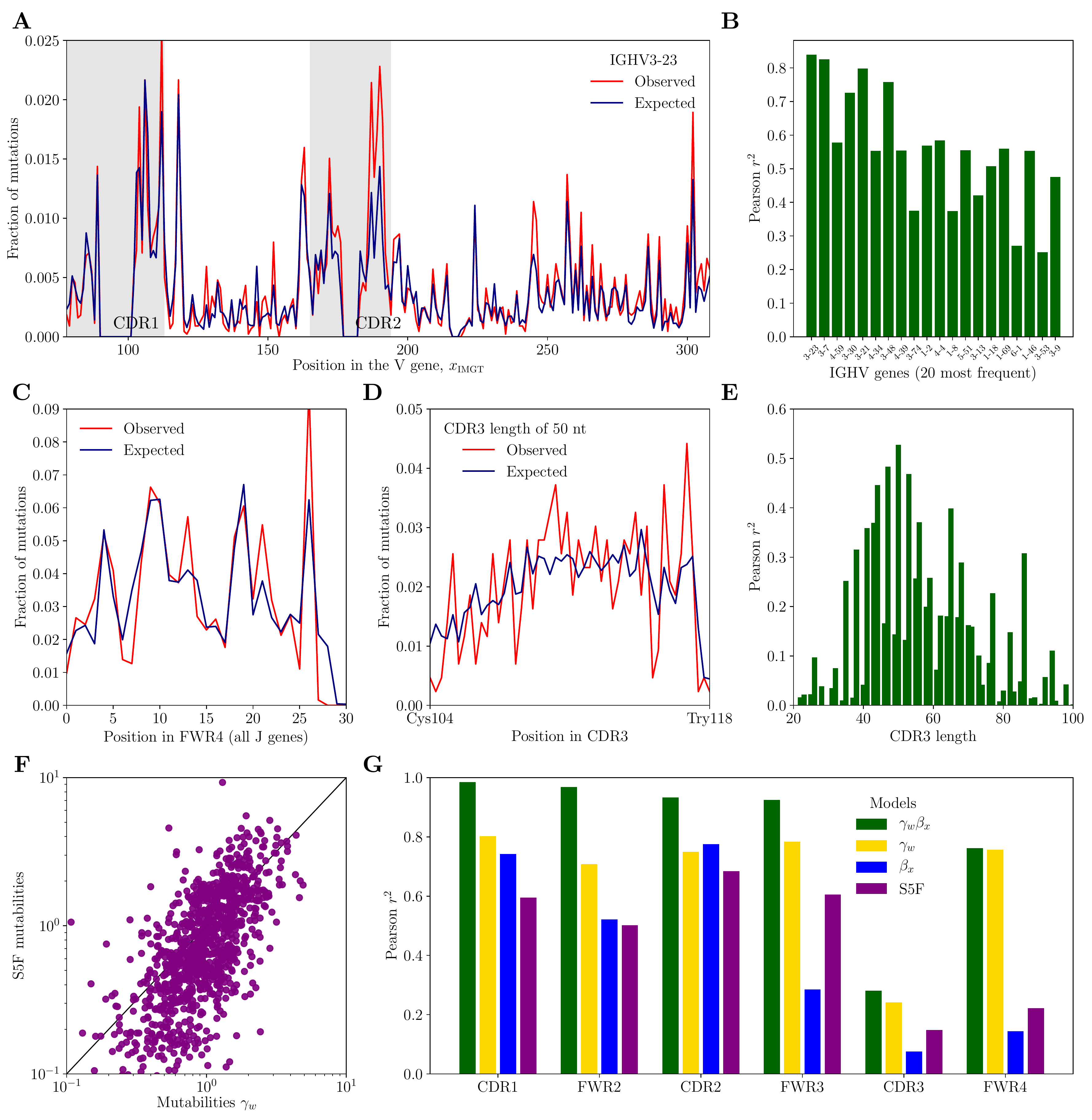}
\end{center}
\caption{
Model performance when model was trained on 2/3 of data, and tested on the remaining 1/3.
A. Example mutation profile in the most common V gene. 
B. Model performance across V genes.
C. Mutation profile in the FWR4 region.
D. Example mutation profile in the CDR3 region for CDR3 length of 50 nts.
E. Model performance across CDR3 lengths.
F. Comparison with the S5F model.
G. Summary of models performance across sequence regions.}
\label{fig:explain_data_21}
\end{figure*}

\begin{figure*}
\begin{center}
  \includegraphics[width=\textwidth]{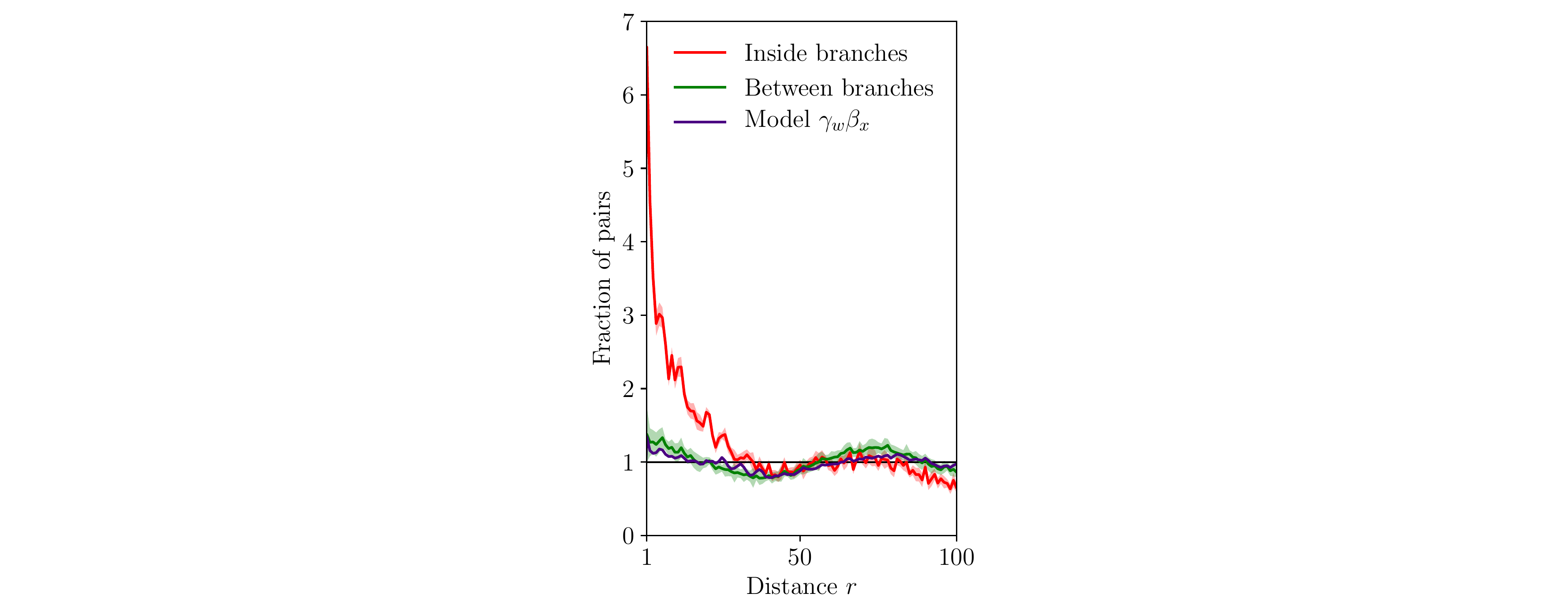}
\end{center}
\caption{Fraction of pairs $n(r)$ of mutations encoded in the same tree branches (red) and pairs of mutations between different branches (green) normalized by the total number of pairs at that distance, $(l-r)/{l \choose 2}$, where $l$ stands for the alignment length. Shaded area corresponds to $95\%$ confidence interval.}
\label{fig:fraction_of_pairs}
\end{figure*}

\begin{figure*}
\begin{center}
  \includegraphics[width=\textwidth]{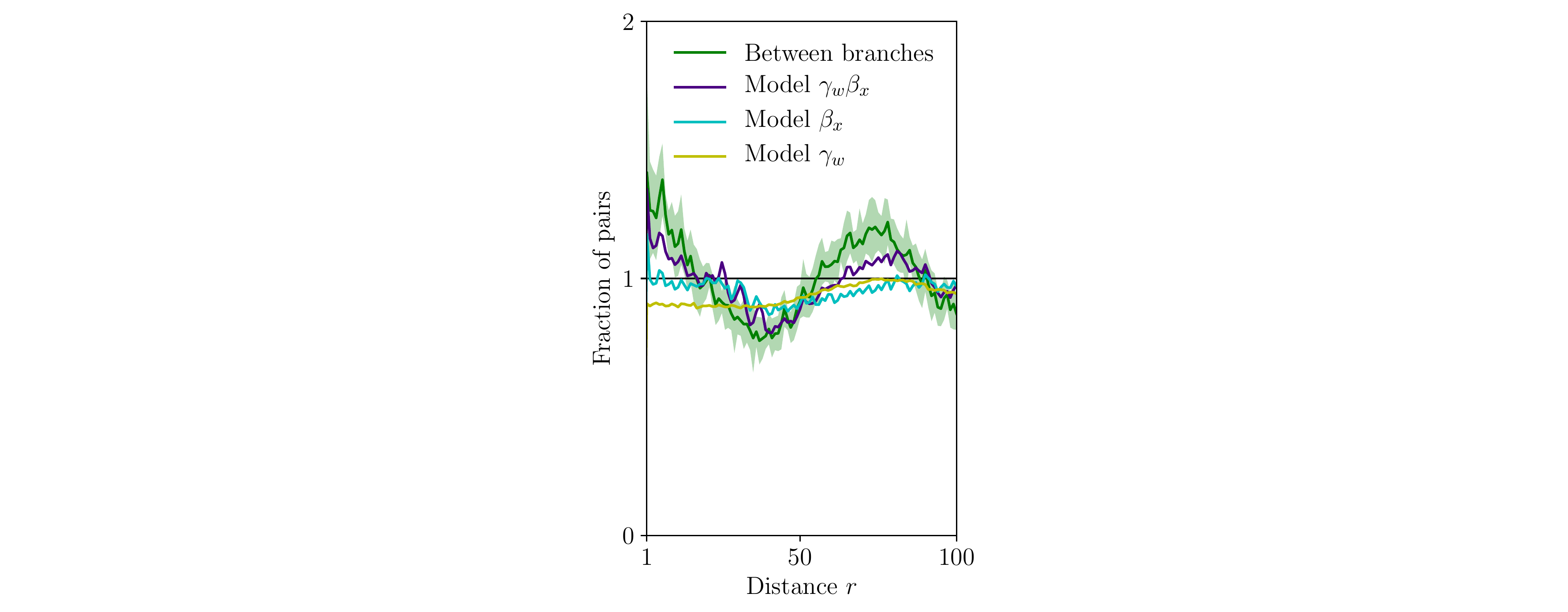}
\end{center}
\caption{Normalized fraction $n(r)$ of pairs of mutations between different branches (green). Context- and position-dependent models prediction $n_m(r)$ for this quantity are presented for the full ($\mu_{s,x}=\gamma_w\beta_x$), purely context-dependent ($\mu_{s,x}=\gamma_w$) and purely position-dependent ($\mu_{s,x}=\beta_x$) models. Shaded area corresponds to $95\%$ confidence interval.}
\label{fig:fraction_of_pairs_models}
\end{figure*}

\begin{figure*}
\begin{center}
  \includegraphics[width=\textwidth]{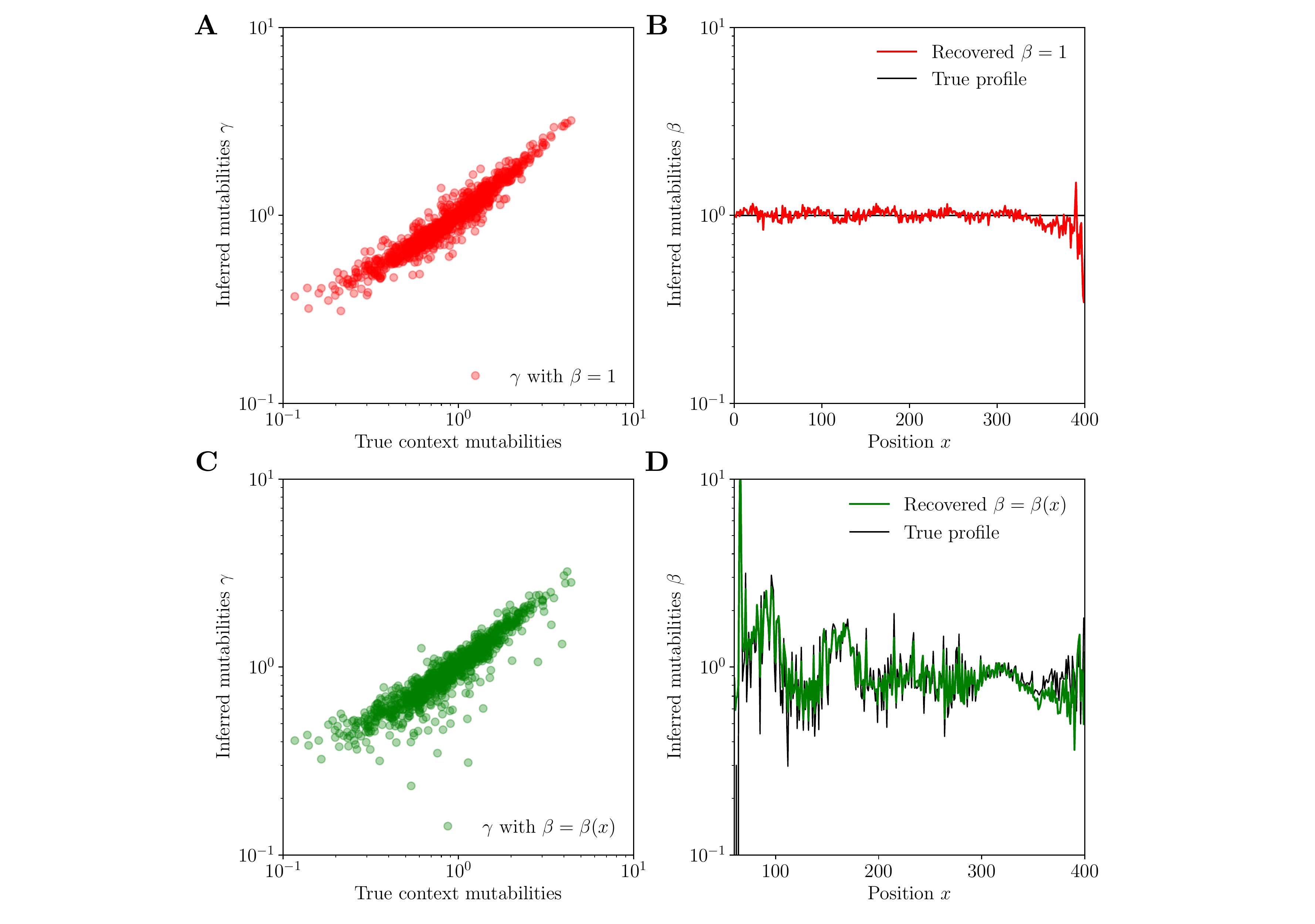}
\end{center}
\caption{Model inference on synthetic data with correlated mutations using true phylogenies. Mutations were drawn according to the co-localization model (\ref{new_toy}) with $\epsilon=5\%$ and $\xi=10$ with data-derived context and position dependent mutabilities. A,C. Inference of context mutabilities $\gamma$. B,D. Inference of position mutabilities $\beta$ for flat (red, A,B) and data-derived (green, C,D) position-dependent profiles.}
\label{fig:reinfer_epsilon}
\end{figure*}

\include{si_table}

\end{document}